\documentclass[preprint,12pt]{elsarticle}

\usepackage{amssymb}

\usepackage{lineno}

\biboptions{sort&compress}

\journal{Astroparticle Physics}

\newcommand{\comment}[1]{}

\begin{document}

\begin{frontmatter}



\title{Asymmetry of the angular distribution of Cherenkov photons of extensive air showers induced by the geomagnetic field}


\author[a1,a2,a3]{P. Homola\corref{cor1}}
\ead{Piotr.Homola@ifj.edu.pl}
\author[a4]{R. Engel}
\author[a1]{H. Wilczy\'nski}
\cortext[cor1]{Corresponding author: Tel.: +48 12 662 8348; fax: +48 12 662 8012.}
\address[a1]{H.~Niewodnicza\'nski Institute of Nuclear Physics, Polish Academy of Sciences, Poland}
\address[a2]{University of Siegen, Germany}
\address[a3]{University of Wuppertal, Germany}
\address[a4]{Institut f\"ur Kernphysik, Karlsruhe Institute of Technology (KIT), Germany}

\begin{abstract}
The angular distribution of Cherenkov light in an air shower is closely linked
to that of the shower electrons and positrons. As charged particles in extensive
air showers are deflected by the magnetic field of the Earth, a deformation of
the angular distribution of the Cherenkov light, that would be approximately
symmetric about the shower axis if no magnetic field were present, is expected.
In this work we study the variation of the Cherenkov light distribution as a
function of the azimuth angle in the plane perpendicular to shower axis. It is
found that the asymmetry induced by the geomagnetic field is most significant for
early stages of shower evolution and for showers arriving almost perpendicular to the vector of the
local geomagnetic field. Furthermore, it is shown that ignoring the azimuthal asymmetry of
Cherenkov light might lead to a significant under- or overestimation of the
Cherenkov  light signal especially at sites where the local geomagnetic field is
strong. Based on CORSIKA simulations, the azimuthal  distribution of Cherenkov
light is parametrized in dependence on the magnetic field component
perpendicular to the shower axis and the local air density. This parametrization  provides
an efficient approximation for estimating the asymmetry of the Cherenkov light
distribution for shower simulation and reconstruction in cosmic ray and
gamma-ray experiments in which the Cherenkov signal of showers with energies above
10$^{14}$~eV is observed.
\end{abstract}


\end{frontmatter}


\section{Introduction}
\label{sec-intro}

Many imaging and non-imaging techniques of observing  air showers are based on
the detection of the abundant number of  photons produced as Cherenkov radiation
of the secondary electrons and positrons in these showers (see, for
example,~\cite{cassidy-casa-blanca97,Antokhonov:2011zz,Berezhnev:2012ys,Tluczykont:2011wq,Voelk:2008fw}).
Cherenkov light also constitutes an important contribution~\cite{Unger:2008uq}
to the optical signal recorded by  fluorescence telescopes built for the
observation of  air showers above $10^{17}$\,eV~\cite{Baltrusaitis:1985mx,Boyer:2002vn,Abraham:2009pm,Tokuno:2012kx}.

With a typical Cherenkov angle of the order of $1^\circ$ in air, the angular
distribution of Cherenkov light around the shower  axis reflects the angular
distribution of the charged particles, mainly electrons and positrons. A proper
estimation of the angular  distribution of the Cherenkov light produced at
various stages of the shower evolution is important for the reconstruction of
the  shower observables and, hence, the parameters of the primary particle.

Already Hillas noticed in his pioneering work on Cherenkov light production in
electromagnetic showers that the energy and angular distributions of electrons
exhibit universality features~\cite{Hillas:1982vn,Hillas:1982wz}. He derived
compact analytic approximations based on the simulation of showers initiated
by photons of $100$\,GeV.
Hadronic showers are subject to much larger fluctuations, limiting the
applicability of universality-based approximations to much higher shower energies.
Only at energies above $\sim 10^{17}$\,eV, the energy, angular, and lateral
distributions of electromagnetic particles in hadronic showers can be efficiently
described by universal functions of shower age and lateral distance in Moli\`ere units, 
e.g.~\cite{Giller:2004fk,Giller:2005qz,Nerling:2006yt,Lipari:2008td,Lafebre:2009en,Ave:2011x1}.

Based on such approximations several parametrizations of the angular
distribution of Cherenkov light have been derived,
see~\cite{Hillas:1982vn,Baltrusaitis:1985mx,Nerling:2006yt,Giller:2009x1}. In
these studies, the angular distribution of Cherenkov photons is considered as
approximately symmetric about the shower axis  and the influence of the local
magnetic field has been neglected. The effects of the geomagnetic field have
been studied so far only for primary photons with energies not larger than
$100$\,GeV~\cite{Elbert83a,chadwick99,commichau08} and a compact parametrization was
derived in Ref.~\cite{Elbert83a}.

The purpose of this work is the quantification of the expected asymmetry of the
Cherenkov light distribution of air showers for a wide range of
primary energies, extending up to the highest cosmic ray energies. Using
CORSIKA~\cite{Heck98a} simulations a parametrization of the angular asymmetry induced by the
geomagnetic field is derived  for air showers from TeV energies up to the
highest energies.  Considering different local magnetic field strengths and
shower  geometries it is discussed under what conditions this asymmetry needs to
be taken into account.

\begin{figure}[htb!]
\begin{center}
\includegraphics[height=12cm,angle=0]{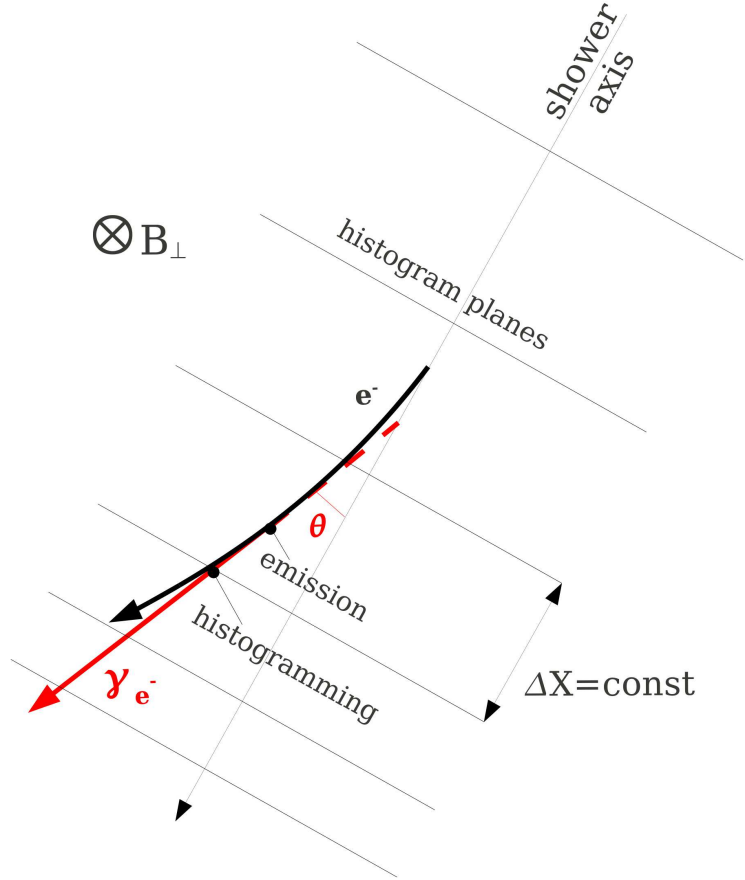}
\caption{Definition of geometric quantities of relevance to this study.
An observer viewing the shower under the angle $\theta$
with respect to the shower axis will see the Cherenkov photons as drawn.
The viewing angle $\theta$ is the angle between the trajectory of a
Cherenkov photon and the shower axis, and
$B_{\perp}$ denotes the component of the geomagnetic field vector
perpendicular to the shower axis. To derive a parametrization the angle
of the Cherenkov photons emitted by shower
electrons is histogrammed at planes perpendicular to the shower axis.
Since we are only interested in the angular distribution of the Cherenkov photons,
the lateral distance of the place of their production relative to the shower
axis is not considered.
}
\label{fig-sim-lay1}
\end{center}
\end{figure}

\begin{figure}[t]
\begin{center}
\includegraphics[height=6cm,angle=0]{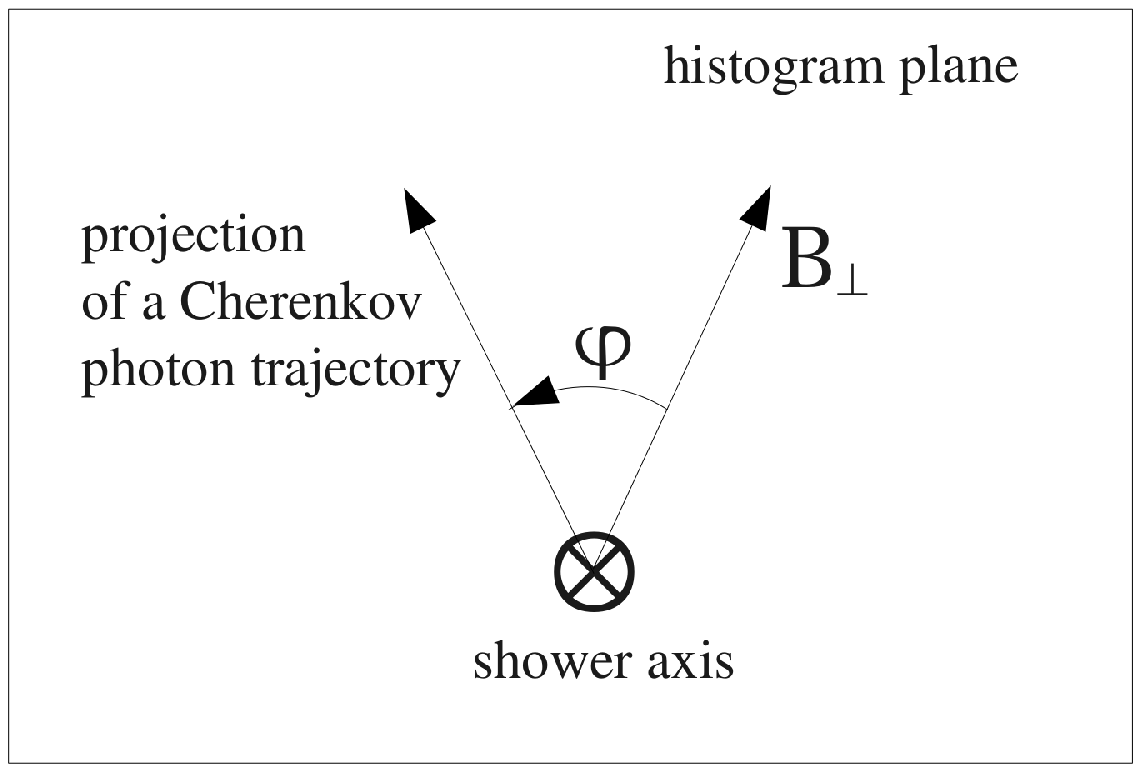}
\caption{The view of the plane in which the Cherenkov photons are histogrammed.
The azimuth angle $\phi$ used within this document is the angle 
between $B_{\perp}$ and the projection of a Cherenkov photon trajectory
onto the plane perpendicular to the shower axis. The angle $\phi$ is measured 
counter-clockwise starting from $B_{\perp}$.} 
\label{fig-sim-lay2}
\end{center}
\end{figure}

Furthermore we show how the asymmetry of the azimuthal distribution of Cherenkov
photons varies for different shower parameters  and geomagnetic conditions. We
replace the commonly used angular distribution of Cherenkov photons, assumed to depend
only on the viewing angle to the shower axis, $F(\theta)$, with a more accurate
function $F(\theta,\phi)$ that also depends on the azimuth angle in the plane perpendicular  to the
shower axis. The definition of the angles is shown in Figs.~\ref{fig-sim-lay1}
and \ref{fig-sim-lay2}.

Throughout this paper we will use the term \textit{viewing angle} for the angle
$\theta$ between the trajectory of a Cherenkov photon and the shower axis (see
Fig.~\ref{fig-sim-lay1}). The azimuth angle $\phi$ used in this study is the
angle between the projection of a Cherenkov photon trajectory onto  the plane
perpendicular to the shower axis, see Fig.~\ref{fig-sim-lay2}, and the projection of the
geomagnetic field vector onto this plane, which will be referred to as $B_{\perp}$.

The Monte Carlo simulations presented throughout this paper were carried out
with CORSIKA~6.970~\cite{Heck98a}. The high energy interactions were processed
with the QGSJET~01 model~\cite{Kalmykov:1997te} and  particles in low energy
($E < 80$\,GeV) were treated with GHEISHA~\cite{Fesefeldt85a}. All the simulations
were performed using the US Standard Atmosphere density
profile~\cite{NASA:1976x1}.  To optimize the computing time we used
the thinning algorithm~\cite{Kobal:2001jx} available in CORSIKA. This algorithm
keeps only a fraction of the secondary particles in a shower below an adjustable
threshold energy, the so-called thinning level.
Only one of the particles of each interaction below this energy threshold is
followed and an appropriate weight is given to it, while
the other particles are dropped. Although the
thinning algorithm introduces additional, artificial fluctuations, it is necessary to apply
it to keep the computing times manageable. For the purpose of this study we
used the thinning level of $10^{-6}$.
  
The Monte Carlo results produced with CORSIKA were processed with COAST~3.01
with the {\it rootrack} option~\cite{ulrich-coast-rootrack}. COAST (COrsika dAta
accesS Tools) is a library of C++ routines providing simple and standardized
access to CORSIKA data. The option {\it rootrack} enables histogramming shower
particles within user-defined planes perpendicular to the shower axis.  For the
purpose of this study a few dedicated modifications were introduced into both
the CORSIKA and COAST codes to enable histogramming of Cherenkov photons. All
the shower simulations used for deriving the parametrization were made with 40
observational levels (planes perpendicular to the shower axis) 
spaced by a constant atmospheric depth interval $\Delta
X=25$\,g/cm$^2$ (see Fig.~\ref{fig-sim-lay1}).  The width of the binning in
$\theta$ is $2^{\circ}$, while the bin width of the azimuth $\phi$ is
$10^{\circ}$.

With use of the above tools we have developed an efficient parametrization that
can be used at an arbitrarily selected experimental site if the local
geomagnetic field vector is known. As an example, the location of the Tunka
experiment~\cite{Berezhnev:2012ys}  (51$^{\circ}$ 48' N, 103$^{\circ}$ 04' E) is
considered in the following. In this experiment a surface array of non-imaging
photon detectors is used to record the Cherenkov light emitted by extensive air
showers of energies between $10^{14}$\,eV and $10^{18}$\,eV.  With the
geomagnetic field being exceptionally strong at the Tunka site ($B=0.6$\,G), it
is well-suited to verify the results of this study experimentally. Even though
the angular distribution of the photons cannot be measured directly with the
Tunka detectors, it can be derived straightforwardly from the measured lateral
distribution of the Cherenkov light at ground.


\section{Cherenkov radiation in extensive air showers}

Cherenkov radiation in air showers is emitted by charged secondaries traveling
with velocity larger than the speed of light in the surrounding medium. There
are three physical parameters important for the overall characteristics of
Cherenkov  light production, namely (a) the threshold energy of the charged
particle above which the Cherenkov radiation is emitted ($E_T$, for altitudes below
$15$\,km  $E_T=30-50$\,MeV), (b) the number of produced photons per unit track
length of a charged particle, and (c) the angles of their emission. All these
quantities depend on the index of refraction $n$ of the medium (i.e.\ air), and $n$
in turn depends on the local density of the air, $n=n(\rho)$. The dependence of $n$ on the
wavelength $\lambda$ is negligible in the typical $\lambda$ range of  interest for
air Cherenkov and fluorescence experiments ($\lambda = 300 \dots
400$\,nm)~\cite{bernloehr08}.

The angular distribution of the Cherenkov light emitted at angles larger than
a few degrees with respect to the shower axis is directly related to the angular
distribution of $e^\pm$ in the shower since the typical Cherenkov angles of less
than $1^\circ$ are negligible at such angular scales. In terms of air shower parameters,
the angular distribution of Cherenkov light should therefore depend mainly on the
angular distribution of charged particles due to multiple scattering, on the
local magnetic field strength perpendicular 
to the trajectory of the particles, and on the air density at the emission point.


\subsection{Universality of electron and Cherenkov photon distributions}
\label{subsec-univ}

There have been many studies of universality features of high-energy air
showers, see, for example, \cite{Giller:2004fk,Giller:2005qz,Nerling:2006yt,Lipari:2008td,Lafebre:2009en,Ave:2011x1}.
In these studies it has been shown that
most distributions of the bulk of the secondary particles can be described by universal
functions depending only on a small number of scaling parameters. Regarding
distributions related to the longitudinal shower evolution, the most important
parameter is the {\it shower age} $s$. The definition of shower age follows from
the analytic treatment of electromagnetic cascades~\cite{Rossi:1941zz} and
applies only to electromagnetic showers. To extend the concept to showers in
general, it is common to use the phenomenological definition~\cite{Hillas:1982wz}
\begin{equation}
s=\frac{3}{1+2X_{\rm max}/X},
\end{equation}
where $X$ is the atmospheric slant depth in g/cm$^2$ and $X_{\rm max}$ is the
slant depth of shower maximum.

\begin{figure}[htb!]
\begin{center}
\includegraphics[height=8cm,angle=0]{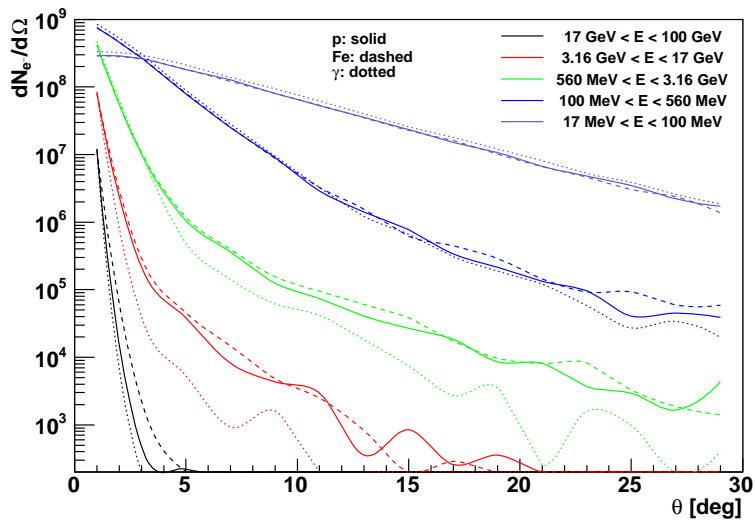}
\caption{
 Angular distribution of electrons in three vertical showers induced by
 a primary proton, iron nucleus, and photon of energies $10^{18}$\,eV 
 for a vanishing geomagnetic field. The distributions are shown at a depth
 corresponding to the maximum shower size (shower age $s=1$).
}
\label{fig-pc-all}
\end{center}
\end{figure}

In the following we only illustrate the universality of
relevant particle distributions in air showers 
by showing the angular distribution of electrons for different energy intervals and
different primary particles, including primary photons (Fig.~\ref{fig-pc-all}). The
simulated distributions were obtained for a vanishing geomagnetic field strength and the
histograms were smoothed to provide a better distinction between the
electrons of different energy ranges.
Each curve corresponds to one shower for each of the primary particles.
As expected, the comparison of the curves in Fig.~\ref{fig-pc-all} shows no significant
differences of the angular distributions of electrons of energies of relevance
to Cherenkov light production. The differences found for the higher energy ranges
are related to shower-to-shower fluctuations and statistical fluctuations due to
the limited number of particles in these histograms. These differences are
unimportant for this study as the low-energy electrons are much more abundant and,
hence, dominate the expected Cherenkov signal.

Averaged over azimuth angle, the angular distribution of electrons
in showers of $10^{17}$\,eV or higher does not
depend significantly on $s$ (in the range of $0.7 < s < 1.2$), $E_0$ nor on
the primary type -- if the primary is a nucleus (see
e.g.~\cite{Giller:2005qz,Nerling:2006yt,Lafebre:2009en}).


The universality of the
electron angular distribution leads to a similar universality of the angular
distribution of Cherenkov photons. Parametrizations of this distribution as a
function of shower age and height can be found e.g.\ in~\cite{Nerling:2006yt}
and \cite{Giller:2009x1}. These parametrizations are very good approximations of
the true angular distribution if the local geomagnetic field component
perpendicular to the shower  axis is small or if one considers distributions
averaged over the azimuthal angle relative to the shower axis.




\subsection{Influence of the geomagnetic field}

Since trajectories of charged particles in air showers are bent by the
geomagnetic field, an azimuthal asymmetry is expected to be present in the angular
particle distribution. This asymmetry leads to an asymmetric
distribution of the Cherenkov light as well as to an asymmetric lateral
distribution of shower particles at ground. The latter is experimentally
well established (see, for example, \cite{Abreu:2011ki}).

The deflection of $e^\pm$ in the geomagnetic field leads to local maxima at two
different azimuth angles in the combined electron and positron
distribution. One maximum is related to $e^{-}$ and the other to $e^{+}$.
Furthermore, due to the larger number of $e^-$ in a shower in comparison to
$e^+$ (see e.g.~\cite{Hillas:1982vn}),
the maximum related to $e^{-}$ is higher than that of $e^{+}$.
In consequence, the azimuthal distribution of
Cherenkov light should be similarly asymmetric.

For investigating this asymmetry with simulated showers and deriving
a parametrization we will follow the original
approach by Hillas~\cite{Hillas:1982vn}. The effective path length $x_m$,
over which the magnetic field typically acts on a
charged particle of energy $E$ and angle $\theta$ wrt.\ the
shower axis, can be approximated by~\cite{Hillas:1982vn}
\begin{equation}
\langle x_m \rangle=\frac{30w^{0.2}}{(1+36\,{\rm MeV}/E)} ({\rm g /cm^2}),
\hspace*{0.5cm}
 w=2(1-\cos\theta)(E/21\,{\rm MeV})^2 .
\label{eq-hillas-xm}
\end{equation}
One obtains $x_m\sim 0.7 X_0$ for $E=100$\,MeV and $\theta=15^\circ$,
using $X_0 = 36$\,g/cm$^2$ for the radiation length in air. 
The magnetic path length $x_m$ increases with $E$ and $\theta$,
and increased $x_m$ results in a larger asymmetry of azimuthal
distributions of charged particles. The expected asymmetry also
increases with the size of the geomagnetic field component
transverse to the direction of the particle momentum, $B_{\perp}$,
i.e.\ the angular distribution will be more elliptical with increasing $B_{\perp}$.

\begin{figure}[htb!]
\begin{center}
\includegraphics[height=8cm,angle=0]{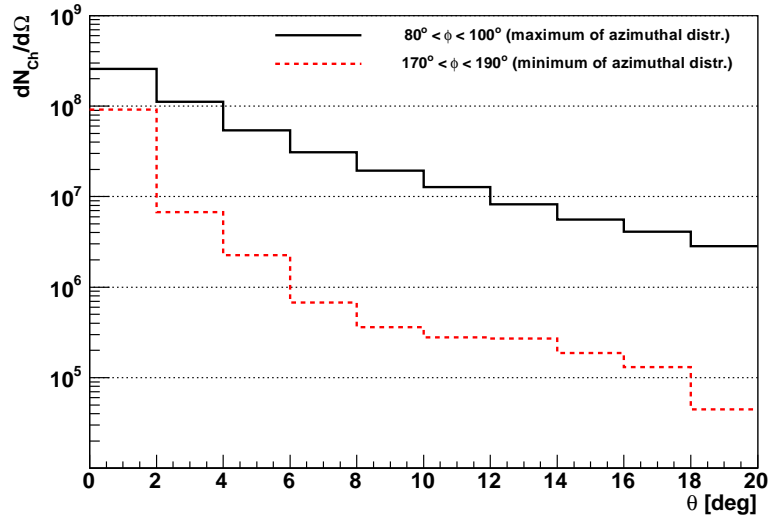}
\caption{Example of a large asymmetry in angular distributions
 of Cherenkov photons. The number of Cherenkov photons is shown
 with respect to the viewing angle for two intervals of the azimuth angle.
 The first azimuth interval, $80^{\circ}<\phi<100^{\circ}$ (solid black),
 corresponds to the viewing angles for which the maximum light intensity
 is observed, while the second interval, $170^{\circ}<\phi<190^{\circ}$ (dashed red),
 is related to the angles for which the light flux reaches its minimum.
 The simulation parameters in this example are: proton as primary particle;
 shower zenith angle $70^{\circ}$; primary energy $10^{15}$\,eV;
 observation level is set at $s=0.8$, corresponding to an altitude of $18-20$\,km;
 transverse component of the geomagnetic field $0.6$\,G
 (compare Fig.~\ref{fig-extreme-hist}).}
\label{fig-cher-domega}
\end{center}
\end{figure}

How large an asymmetry of the angular distribution of Cherenkov light can arise
is illustrated by an example shown in Fig.~\ref{fig-cher-domega}, where the
number of  Cherenkov photons is integrated over two different intervals of
azimuth $\phi$ and shown with respect to the viewing angle $\theta$. The two
histograms show the angular distribution of the Cherenkov light for a proton
shower of energy $E_0=10^{15}$\,eV arriving at a zenith angle  of $70^{\circ}$.
The light emission was observed before the shower reached its maximum
development (shower age $s=0.8$) 
assuming a geomagnetic field component of $B_{\perp}=0.6$\,G.  
The parameters were chosen to maximize the expected
asymmetry. The difference in Cherenkov light intensity emitted from the selected
regions of the shower observed at different viewing directions exceeds an order
of magnitude. For weaker magnetic fields the differences between the maximum and
minimum intensities are less pronounced, but still clearly visible.

\begin{figure}[htb!]
\begin{center}
\includegraphics[height=8.0cm,angle=0]{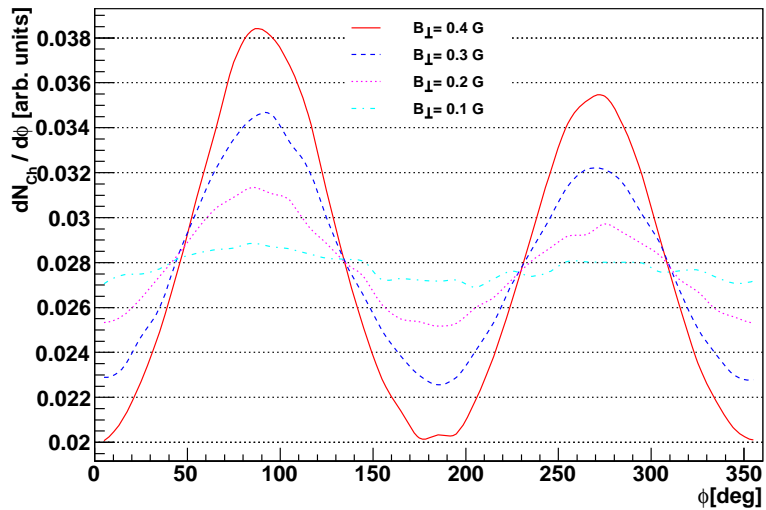}
\caption{Cherenkov photons in air showers vs.\ the azimuth angle for $B_{\perp}= 0.1, 0.2, 0.3, 0.4$\,G. The simulation parameters 
in these examples are: proton primaries, shower zenith angle 
$= 60^{\circ}$, primary energy $= 10^{19}$\,eV, observation level at $s=1$ 
(shower maximum, corresponding to an altitude of $8-10$\,km), viewing
angle $8^{\circ}<\theta<10^{\circ}$.}
\label{fig-elec-and-chph-btr}
\end{center}
\end{figure}

Some representative examples of azimuthal distributions of Cherenkov photons for
fixed viewing angle and varying $B_{\perp}$ are shown in
Fig.~\ref{fig-elec-and-chph-btr}. The left peak of the Cherenkov distributions is
related to deflected electrons and the right one to positrons.
The electron peak is higher than the positron one due to the
$20-30$\% excess of electrons in the shower disk (see
e.g.~\cite{Hillas:1982vn,Bergmann:2006yz}). The peak heights of the
distributions shown in Fig.~\ref{fig-elec-and-chph-btr}
and the asymmetry in Fig.~\ref{fig-cher-domega} are different because of
different altitudes at which the Cherenkov light is produced.
A higher air density leads to shorter trajectories of electrons
and positrons and correspondingly less geomagnetic deflection.

Figs.~\ref{fig-cher-domega} and \ref{fig-elec-and-chph-btr} confirm that a
parametrization properly describing the angular distribution of Cherenkov light
should include a term for the dependence on the azimuthal angle $\phi$ in the
plane perpendicular to the shower axis. Accounting only for the dependence on
the viewing angle $\theta$ may not be a good approximation in all cases.


\section{Parametrization of azimuthal Cherenkov light distribution}
\label{sec-param}

Universality of the underlying electron and positron distributions implies
that the azimuthal profiles of Cherenkov photons  can also be described by a
universal function if expressed in suitable variables. Following Elbert {\it
et~al.}~\cite{Elbert83a},  we make the ansatz that the asymmetry of the
azimuthal profile of Cherenkov photons can be approximated by a function of the
azimuth angle $\phi$, the viewing angle $\theta$, and the parameter $a$ defined
as
\begin{equation}
 a = B_{\perp}/\rho,
\end{equation}
where $\rho$ is the local density of air. A dependence on the air density is
expected due to the interplay between the magnetic track length $x_m$, see
Eq.~(\ref{eq-hillas-xm}), and multiple Coulomb scattering. In addition the
energy threshold for Cherenkov emission depends on the air density through the
density dependence of the refractive index $n=n(\rho)$. Applying the
parametrization of $n(\rho)$ as implemented in CORSIKA we neglect the
variation of $n$ with wavelength.

In the following we consider proton, nuclei, and photon induced showers
in the energy range $10^{15}\,{\rm eV} < E_0 < 10^{17}$\,eV and focus on
shower ages $0.7 < s < 1.2$, which are of relevance to shower detection. 

For the purpose of deriving a parametrization of the azimuthal distributions of
Cherenkov photons, we simulated proton-induced  showers with three primary
energies ($10^{15}$, $10^{16}$ and $10^{17}$\,eV) arriving at two different
zenith angles ($58^{\circ}$  and $70^{\circ}$) for different values of
$B_{\perp}$ (0.2, 0.3, 0.4, 0.5, and 0.6\,G). These simulation parameters were
selected to  cover a reasonably wide range of the parameter $a$ and to ensure
acceptable computing time for producing the shower library.

\begin{figure}[htb!]
\begin{center}
\includegraphics[width=13cm,angle=0]{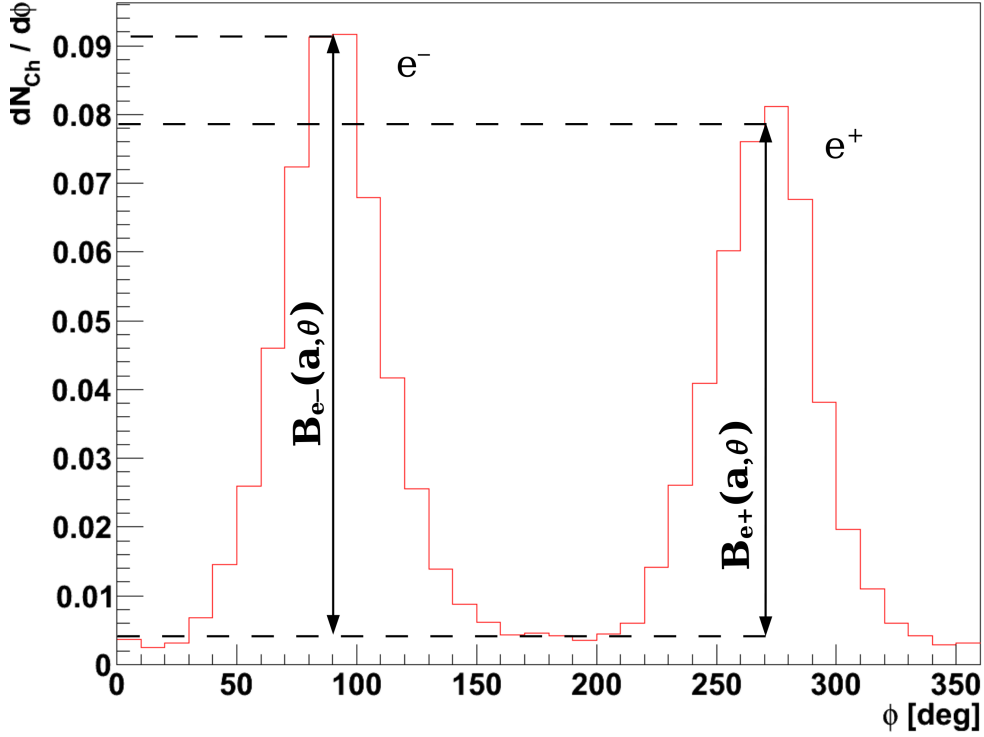}
\caption{Example histogram of Cherenkov photons from one of the simulations
used for deriving the parametrization. The histogram was obtained for a proton
primary of $10^{15}$\,eV arriving at a zenith angle of $70^{\circ}$,
with $B_{\perp}$=0.6\,G. Only photons emitted at an
shower age $s\approx0.8$ and viewing angle $14^{\circ}<\theta<16^{\circ}$ are shown. 
In this case $a=2.9$\,G\,m$^3$\,kg$^{-1}$. See text for details
concerning the peak heights $B_{e+}$ and $B_{e-}$.}
\label{fig-extreme-hist}
\end{center}
\end{figure}

Of the 40
observation levels, the first of them is always located at the top of the
atmosphere which corresponds to a vertical height of 100\,km in CORSIKA.  To obtain an
azimuthal distribution of the emitted Cherenkov radiation for segments of the
shower, we fill photons emitted between the two adjacent observational
planes in histograms for different viewing angles $\theta$. For each combination of primary
energy,  arrival direction, and $B_{\perp}$, a simulation  run for a single
primary particle was repeated 10 times with different random seed
initialization. All the resulting azimuthal  distributions of Cherenkov photons
shown within this paper are normalized to 1.  An example of such a histogram of
Cherenkov photons is shown in Fig.~\ref{fig-extreme-hist}.

\begin{figure}[htb!]
\begin{center}
\includegraphics[width=1.0\textwidth,angle=0]{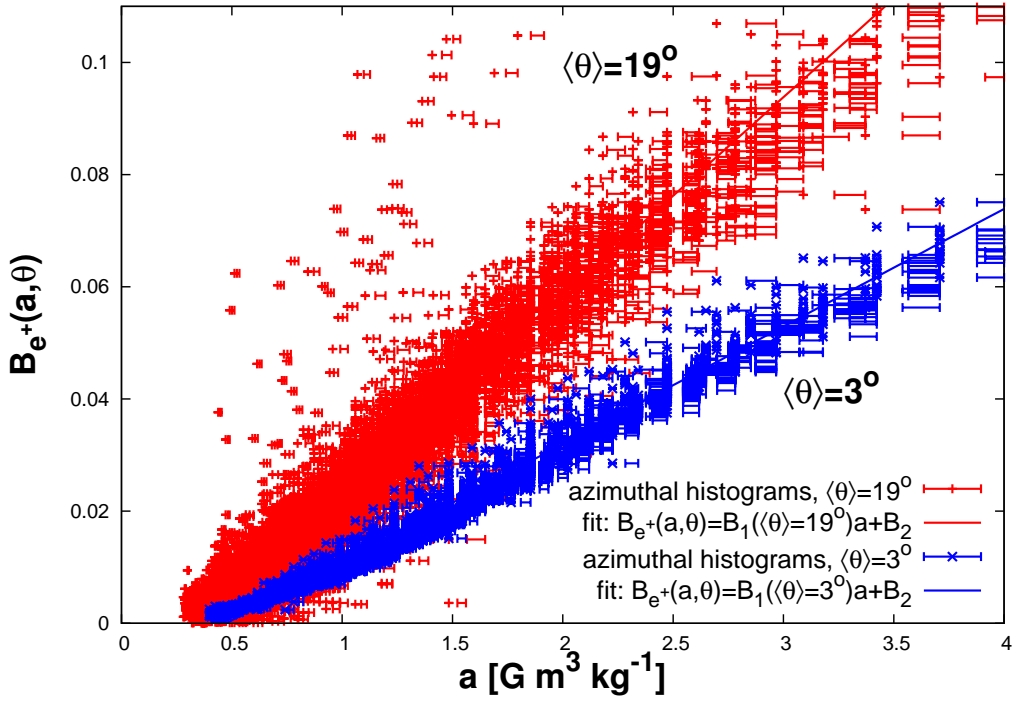}
\caption{Dependence of peak height $B_{e^{+}}(a,\theta)$ of normalized azimuthal
Cherenkov profiles on asymmetry parameter $a$ 
for different simulation parameters and two different angles $\theta$.
The symbols with represent the simulation results used for the fit.
The error bars indicate the uncertainty in $a$ resulting from the distance
of the different planes for observation.
For other details and parameters of the fitted functions, see text.
}
\vspace{-0.2cm}
\label{fig-normalized-amplitude}
\end{center}
\end{figure}

In the next step of the analysis we process each of the selected histograms to
find the heights of the peaks related to  positrons ($B_{e^{+}}$) and electrons
($B_{e^{-}}$) (see Fig.~\ref{fig-extreme-hist}). The heights are read as the
average ``maximum'' values of the adjacent azimuth bin contents at
$80^\circ<\phi\leq90^\circ$ and $90^\circ<\phi\leq100^\circ$ for $B_{e^{-}}$ and
at $260^\circ<\phi\leq270^\circ$ and $270^\circ<\phi\leq280^\circ$ for
$B_{e^{+}}$, subtracting the average ``minimum'' bin contents at  $\phi=0^\circ$
(bins $350^\circ<\phi\leq360^\circ$ and $0^\circ<\phi\leq10^\circ$) and at
$\phi=180^\circ$  (bins $170^\circ<\phi\leq180^\circ$ and
$180^\circ<\phi\leq190^\circ$), respectively. The values of $B_{e^{+}}$ and
$B_{e^{-}}$  depend on $a$ and $\theta$. An example of this dependence is shown
in Fig.~\ref{fig-normalized-amplitude}, where $B_{e^{+}}$ is  plotted as a
function of $a$ for 2 different viewing angles $\theta$. For each histogram of
Cherenkov photons an average value  of $a$ is calculated based on the
atmospheric air density $\rho$ at the two levels determining the histogram, with
interpolation  of $\rho$. The maximum uncertainty of this interpolation is
reflected in the horizontal error bars. Fig.~\ref{fig-normalized-amplitude}
indicates that the azimuthal asymmetry increases with $a$ and $\theta$.

The increase of $B_{e^+}$ with $a$ means that the asymmetry is larger for
stronger $B_{\perp}$ or for smaller air density.  A stronger  $B_{\perp}$ leads
to a stronger deflection of the trajectories of charged particles, hence, the
asymmetry of the  azimuthal profile of Cherenkov photons is more pronounced than
in case of weaker $B_{\perp}$. At the same time the mean free  path of electrons
over which they are being deflected increases with decreasing air density, which
also causes an increase of  the azimuthal asymmetry of Cherenkov photons.
Finally, the increase of this asymmetry with $\theta$ is related to the number
of electrons of different energies varying with $\theta$. At larger viewing
angles one expects more low-energy electrons. Electrons of low energy
are deflected by the geomagnetic field more efficiently than those of 
higher energies.

From the experimental point of view it is important to keep in mind that, for
large viewing angles ($\theta > 20^\circ$), the intensity of the  Cherenkov
light emitted from the shower particles will become smaller than the Cherenkov light emitted at smaller
angles in earlier stages of the shower that is later scattered in the atmosphere
to large angles. An investigation of the viewing angle at which the transition
between the dominance of direct Cherenkov light and scattered Cherenkov light
takes place depends on shower parameters and the observation conditions, and is
beyond the scope of this work.

\begin{figure}[t]
\begin{center}
\includegraphics[width=13cm,angle=0]{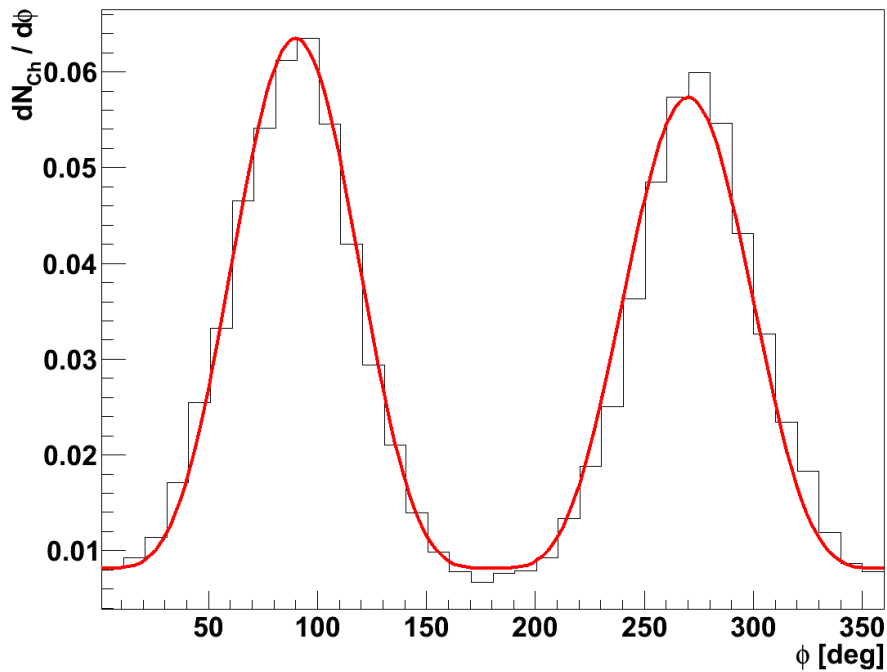}
\caption{Comparison between an arbitrarily selected azimuthal
histogram of Cherenkov photons and the
parametrization~(\ref{eq-new-par}), shown as solid line.
The histogram was obtained for an iron primary of energy $10^{17}$\,eV,
arriving at a zenith angle of $63^{\circ}$,
with $B_{\perp}=0.53$\,G (corresponding to $a=1.96$\,G\,m$^3$\,kg$^{-1}$)
at viewing angle $10^{\circ}<\theta<12^{\circ}$. 
The shower age was approximately $0.75$.
}
\label{fig-test-fit-fe}
\end{center}
\end{figure}

The numerically found values of $B_{e^{+}}$ and $B_{e^{-}}$ can be described
by a linear function of the parameter $a$
\begin{eqnarray}
& &B_{e^{+}/e^{-}}(a,\theta)=B_1(\theta)a+B{_2} \nonumber\\
& &B_1(\theta)=B_{11}\theta^{2}+B_{12}\theta+B_{13},
\end{eqnarray}
where
\begin{eqnarray}
& &B_{2}(e^{+}) = -0.0093 \pm 0.0004 \nonumber\\
& &B_{11}(e^{+}) = (-6.2 \pm 0.5) \times 10^{-5} {\rm deg}^{-2} \nonumber\\
& &B_{12}(e^{+}) = (0.0022 \pm 0.0001) {\rm deg}^{-1} \nonumber\\
& &B_{13}(e^{+}) = 0.0136 \pm 0.0005,
\end{eqnarray}
and
\begin{eqnarray}
& &B_{2}(e^{-}) = -0.0084 \pm 0.0005 \nonumber\\
& &B_{11}(e^{-}) = (-8.5 \pm 0.5) \times 10^{-5} {\rm deg}^{-2} \nonumber\\
& &B_{12}(e^{-}) = (0.0028 \pm 0.0001) {\rm deg}^{-1} \nonumber\\
& &B_{13}(e^{-}) = 0.0124 \pm 0.0005.
\end{eqnarray}
The normalized distribution can be approximated by
\begin{equation}
\label{eq-new-par}
F(a,\theta,\phi)=A+F_{S}(a,\theta)[1+F_{A}(a,\theta)\sin\phi]\,\sin^{4}\phi ,
\end{equation}
where $A=1/(2\pi)-3F_{S}/8$ is the normalization constant and coefficients $F_{S}(a,\theta)$ and $F_{A}(a,\theta)$ are 
given by the heights of the distribution peaks $B_{e^{+}}$ and $B_{e^{-}}$
\begin{eqnarray}
& &F_{S}(a,\theta)=[B_{e^{-}}(a,\theta)+B_{e^{+}}(a,\theta)]/2,\nonumber\\
& &F_{A}(a,\theta)=[B_{e^{-}}(a,\theta)-B_{e^{+}}(a,\theta)]/[B_{e^{-}}(a,\theta)+B_{e^{+}}(a,\theta)].
\end{eqnarray}
An example of a shape of the function~(\ref{eq-new-par}) defined above,
for a set of arbitrary parameters, is presented in Fig.~\ref{fig-test-fit-fe}
in comparison to one shower histogram of Cherenkov photons. The typical 
shape of the azimuthal distribution of Cherenkov photons, including the excess of electrons in
comparison to positrons (i.e.\ the increased height of the ``electron'' peak) is well reproduced.

As mentioned earlier, the density of Cherenkov photons decreases rapidly with
increasing $\theta$. Therefore the quality  of the azimuthal histograms at large
viewing angles $\theta$ is limited by statistics. Also for small angles $\theta$
(near  the shower axis), where the phase space due to the small solid angle 
is substantially reduced, the
statistics of Cherenkov photons is insufficient to obtain good quality
histograms.  As a consequence, the range of the viewing  angles covered by the
presented analysis is limited to $2^{\circ}<\theta<20^{\circ}$.

\begin{figure}[t]
\begin{center}
\includegraphics[height=8cm,angle=0]{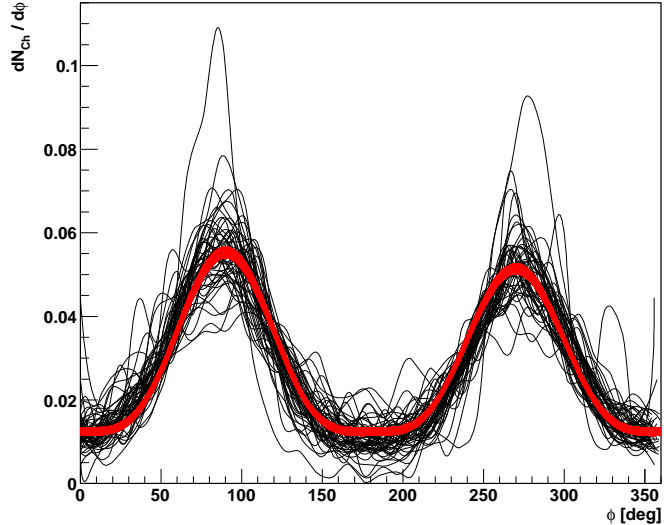}
\caption{Example of shower-to-shower fluctuations of azimuthal
distributions of Cherenkov photons in air showers initiated by
primary protons of $10^{14}$\,eV. The thin black lines represent 50 individual 
air showers and are compared to the parametrization~(\ref{eq-new-par})
for $a=(1.5 \pm 3\%)$\,G\,m$^3$\,kg$^{-1}$ and $\langle\theta\rangle=
19^{\circ}$ (thick red lines).
}
\label{fig-s2s-4}
\end{center}
\end{figure}

The parametrization was tested with a set of reference showers generated for the
geomagnetic location of the Tunka Cherenkov  array. Performing a $\chi^2$ test
it is found that about $90$\% of all simulated histograms deviate less than
$30$\% from the predicted asymmetry for photon-induced showers of primary
energies $E_{0}\geq 10^{14}$\,eV, for  showers initiated by protons with
$E_{0}\geq 10^{15}$\,eV, and for iron-initiated showers with $E_{0}\geq
10^{16}$\,eV. The   deviation of a histogram is considered to be less than
$30$\% if the $\chi^2$, calculated by assuming a $30$\% uncertainty for   each
interval of the histogram, is less than unity per degree of freedom. For
example, the comparison of 50 histograms of proton-induced showers of
$10^{14}$\,eV with the parametrization is shown  in Fig.~\ref{fig-s2s-4}.
At this low energy, fluctuations lead to significant deviations of individual
showers from the parametrization that, by construction, only reproduces
the mean distribution.

\begin{figure}[htb!]
\begin{center}
\includegraphics[height=8cm,angle=0]{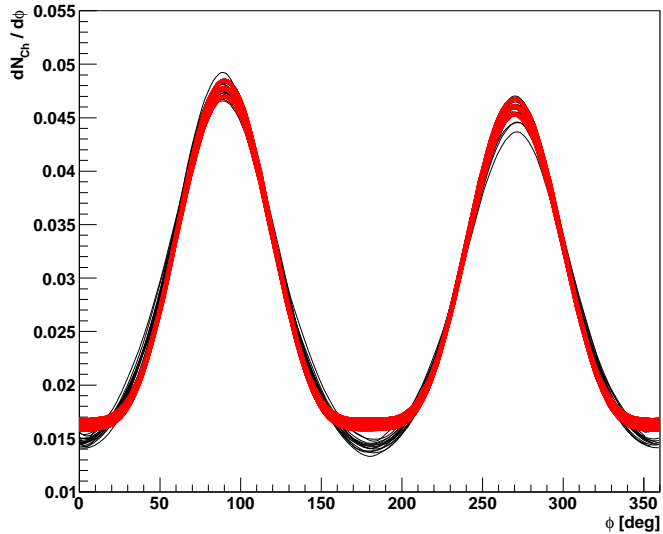}
\caption{Example illustrating the shower-to-shower fluctuations
of azimuthal distributions of Cherenkov
photons in air showers initiated by a primary protons of 
$10^{17}$\,eV. The thin black lines are 9 individual air showers 
that are compared to the corresponding parametrizations (thick red lines). The 
parameters of this comparison are $a=(2.0 \pm 3\%)$\,Gm$^3$kg$^{-1}$
and $\langle\theta\rangle=3^{\circ}$.
}
\label{fig-s2s-6}
\end{center}
\end{figure}

The shower-to-shower fluctuations are rapidly decreasing with increasing energy.
This improves the overall description of the azimuthal Cherenkov asymmetry, as
can be seen in Fig.~\ref{fig-s2s-6} where proton showers of $10^{17}$\,eV are
shown. Deviations between the simulation results and the parametrization are
found for the minima of the distribution.  It has not been attempted to improve the 
description in the regions of the minima as one would have to introduce further
parameters in Eq.~(\ref{eq-new-par}).

\begin{table}[htb!]
\begin{center}
\caption {Examples of values of the parameter $a$ for selected experimental sites, assuming $B_\perp=|\vec B|$. }
\vspace{0.5cm}
{%
\newcommand{\mc}[3]{\multicolumn{#1}{#2}{#3}}
\begin{center}
\begin{tabular}{llllll}
\hline \hline
experiment & \mc{4}{c}{a [Gm$^3$kg$^{-1}$]}\\
    & \mc{1}{c}{10 km a.s.l.} & \mc{1}{c}{8 km a.s.l.} & \mc{1}{c}{5 km a.s.l.} & \mc{1}{c}{3 km a.s.l.}\\
\hline
Auger \cite{Abraham:2009pm} &  \mc{1}{c}{0.57} & \mc{1}{c}{0.45} & \mc{1}{c}{0.33} & \mc{1}{c}{0.27}\\
H.E.S.S. \cite{hess} & \mc{1}{c}{0.67} & \mc{1}{c}{0.53} & \mc{1}{c}{0.39} & \mc{1}{c}{0.32}\\
Telescope Array \cite{Tokuno:2012kx} & \mc{1}{c}{1.20} & \mc{1}{c}{0.96} & \mc{1}{c}{0.70} & \mc{1}{c}{0.57}\\
Tunka \cite{Antokhonov:2011zz} & \mc{1}{c}{1.42} & \mc{1}{c}{1.13} & \mc{1}{c}{0.82} & \mc{1}{c}{0.67}\\
\hline \hline
\end{tabular}
\end{center}
}%
\label{table-a}
\end{center}
\end{table}

In this study, the statistics of histograms are sufficient to confirm the
general applicability of the parametrization (\ref{eq-new-par}) in the range
$0.3$\,G\,m$^3$\,kg$^{-1}<a<2.5$\,G\,m$^3$\,kg$^{-1}$. This range covers the
expected values of $a$ for air shower detectors and air Cherenkov telescopes, see Tab.~\ref{table-a}. 
Asymmetric profiles at small $a$ are less interesting
because they can hardly be distinguished from a flat distribution in azimuth.

It is important to note that the precision of the parametrization~(\ref{eq-new-par}),
although satisfactory within the checked range of  the parameter $a$, decreases
with increasing $a$.  Further studies will be needed to extend the
parametrization to showers of  low energy in the presence of large $B_{\perp}$.
These showers reach their maximum at higher altitudes (smaller $\rho$) than
considered here. It is expected that the peaks in the azimuthal
distributions will be narrower under these conditions
than the ones predicted by Eq.~(\ref{eq-new-par}).


\section{Summary and outlook}

In this work a detailed parametrization of the azimuthal asymmetry of the
angular distribution of Cherenkov photons in  air showers of energies above
10$^{14}$ has been derived (Eq.~\ref{eq-new-par}). The parametrization can be applied
to estimate the asymmetry for geomagnetic field strengths and air densities,
expressed by the parameter $a$, in the range  $0.3$\,G\,m$^3$\,kg$^{-1}<a<2.5$\,G\,m$^3$\,kg$^{-1}$.
The wide range in $a$ ensures that an azimuthal asymmetry of Cherenkov
photons can be predicted well for the majority of showers of relevant
experiments.

The result shows that the asymmetry of the azimuthal Cherenkov distribution
caused by the geomagnetic field is expected to be  significant when the air
density is small and the local magnetic field component transverse to the shower
axis is large (large $a$).   The asymmetry increases also with the angle
between the viewing direction and the shower axis. Ignoring the asymmetry of
the azimuthal  distribution of Cherenkov photons can lead to under- or
overestimation of the Cherenkov component in the shower light profiles by up
to one order of magnitude in the most extreme case.

To illustrate the expected asymmetry for optical air shower observations we consider the ratio
$({\rm d}N_{\rm ch}/{\rm d}\phi)_{\phi=90^{\circ}}/({\rm d}N_{\rm ch}/{\rm d}\phi)_{\phi=180^{\circ}}$
for proton showers of $10^{18.5}$eV, arriving at the site of the Pierre
Auger Observatory~\cite{Abraham:2009pm} from
geographical South at zenith angles of $30^{\circ}$ and $60^{\circ}$. If one
looks at the depth of maximum of these showers at a viewing angle of
$15^{\circ}$ the expected asymmetries of the azimuthal Cherenkov light
distributions are 1.04 and 1.15, respectively.  For $s=0.8$ the asymmetries are
even larger: 1.20 and 1.39. For the Telescope Array site, where the local
geomagnetic field is twice as strong as that at the Auger site, the expected
asymmetries for showers seen at their maxima  at a viewing angle of $15^{\circ}$
and  arriving at the same zenith angles and from geographical North are 1.45 for
$30^{\circ}$ and 2.45 for $60^{\circ}$. The respective  asymmetries seen at
$s=0.8$ are 1.71 and 3.72.

The results of this work are important for cosmic ray experiments in which the
Cherenkov signal of showers is recorded  and the local geomagnetic field is not
very weak. As the parametrization formulas derived in this paper are applicable
also for  photon air showers of primary energies down to $100$\,TeV, they could
also be used in the analysis of data of ground-based  gamma-ray detectors.


\section*{Acknowledgements}
This work was partially supported in Poland by the
Polish Ministry of Science and Higher Education
under grant No.~NN~202 2072 38 and by the National Science Centre, grant No.~2013/08/M/ST9/00728,
and in Germany by the DAAD, project ID 50725595,
the Helmholtz Alliance for Astroparticle Physics
and the BMBF Verbundforschung Astroteilchenphysik.










\bibliographystyle{unsrt-mod-notitle}
\bibliography{local,references}












\end{document}